\documentclass[aps,reprint,nofootinbib,twocolumn,superscriptaddress,longbibliography]{revtex4-2}
\usepackage{bbold}
\usepackage{physics}
\usepackage{comment}
\usepackage{natbib}
\usepackage{slashed}

\usepackage[english]{babel}
\usepackage{graphicx,epstopdf}% Include figure files
\usepackage{amssymb,amsmath}
\usepackage{dcolumn}% Align table columns on decimal point
\usepackage{bm}% bold math
\usepackage{color}
\usepackage{enumerate}
\usepackage[colorlinks=true,citecolor=blue]{hyperref}
\hypersetup{colorlinks=true,citecolor=blue,linkcolor=blue,urlcolor=blue}
\usepackage{soul}
\usepackage{tabularx}
\usepackage[dvipsnames]{xcolor}
\usepackage{appendix}

\begin{document}

\preprint{APS/123-QED}

\title{Thermoelectric effects in two-dimensional topological insulators}% Force line breaks with \\

\author{Z.Z. Alisultanov}
\email{zaur0102@gmail.com}
\affiliation{Abrikosov Center for Theoretical Physics, MIPT, Dolgoprudnyi, Moscow 141701, Russia}
\affiliation{Institute of Physics of DFRS, Russian Academy of Sciences, Makhachkala, 367015, Russia}

\author{E.G. Idrisov}
%\email{edvin.idrisov@uaeu.ac.ae}
\affiliation{Department of Physics, United Arab Emirates University, P.O. Box 15551 Al-Ain, United Arab Emirates}

\author{A.V. Kavokin}
\email{kavokinalexey@gmail.com}
\affiliation{School of Science, Westlake University, Hangzhou 310024, Zhejiang Province, China}
\affiliation{Abrikosov Center for Theoretical Physics, MIPT, Dolgoprudnyi, Moscow 141701, Russia}


\begin{abstract}  
We explore the nontrivial thermoelectric properties of two-dimensional topological systems. For the Chern insulator, we show that the Seebeck coefficient is fully determined by the Kelvin formula, while the Nernst coefficient vanishes. For a two-dimensional electron gas with Rashba spin-orbit interactions we reveal how the Berry curvature affects the thermoelectric coefficients, and derive the Mott-like equation for thermopower. We predict a strong variation of the thermopower of a two-dimensional topological insulator with time-reversal symmetry in the ballistic and dissipative regimes. The Kelvin formula applies in the ballistic regime, while the Mott formula holds in the dissipative regime. Importantly, in a system with trapezoidal geometry, the combination of ballistic and dissipative regimes leads to the anomalous Nernst effect. Finally, we analyze a two-dimensional Anderson insulator, where edge modes show distinct temperature behavior of the Seebeck coefficient near the weak localization-strong localization transition temperatures. In the trivial phase, the thermopower exhibits a strong power law temperature dependence, while in the topological phase both power law and exponential dependences coexist.
\end{abstract}

\maketitle

%\tableofcontents

\section{Introduction}
The study of electric and thermal transport in materials with 
non-trivial band topology continues to be a prominent area of research in modern condensed matter physics~\cite{Hasan,Qi,Volovik_11,Armitage,Volovik_19,Anirban,Tokura,Jin1,Jin2,Park1,Onishi}. A special emphasize is placed on the analysis of thermoelectric effects in such systems, where dissipation may occur~\cite{Wang01,Lee04,Xiao06,ivanov2018thermoelectric,Vargiamidis,Fujikawa}. Shortly after their experimental realization,  these materials were recognized as promising candidates for the observation of nontrivial thermoelectric effects~\cite{Xu,Fu20}. Namely, the large thermoelectric values have been predicted and experimentally observed in gapless topological materials, such as Weyl and Dirac semimetals~\cite{Lundgren,nakatsuji2015large, ikhlas2017large}. Additionally, thermoelectric measurements have been proposed as an effective method for investigating exotic properties of Weyl semimetals, such as the chiral anomaly and Fermi-arc surface states~\cite{Jia, Liang2013}. Furthermore, there is a strong interest in studying topological insulator heterostructures that are expected to exhibit unique transport properties~\cite{Fan,Kandala,Hesjedal,Chong,Tokura,Eremeev,Liu10,Alisultanov_multilayer}. A key motivation for studying such multilayer topological systems is that these structures facilitate the observation of phenomena related to quantum geometry~\cite{Törmä,Gao,alisultanov2024disorder}. It is presumed that this concept serves as a general and foundational approach to topological condensed matter physics, and, in particular, the quantum geometry manifests itself in anomalous nonlinear thermoelectric effects~\cite{Varshney}.

Recent studies have explored thermoelectric phenomena in one-dimensional and two-dimensional ballistic systems~\cite{Kavokin1,Kavokin2}. These works demonstrate that in one-dimensional systems with dimensional quantization, the Kelvin formula must be used, while the Mott formula is not applicable~\cite{Kavokin1}. In ballistic graphene-like systems, the Kelvin formula also cannot be applied due to a non-vanishing anomalous correction~\cite{Kavokin2}. This anomaly arises because equilibrium in a ballistic system is not reached due to the lack of thermalization. Importantly, the zero total current condition imposed by the broken circuit geometry may be fulfilled in the presence of non-damped counter-propagating flows of charge carriers.

In this context, the present study of thermoelectric properties of two-dimensional topological materials is a natural step forward. In this work we show that in the presence of T-symmetry (topological insulators of class AII), the Seebeck coefficient is exactly determined by the Kelvin formula, as transport occurs through ballistic chiral edge modes. In contrast, for Chern insulators, the Mott-type formula applies. We then incorporate inelastic scattering of edge modes into the model and study a trapezoidal-shaped TI, where one base is shorter than the scattering length (that is the condition of ballistic regime) and the other one is longer than the scattering length (that corresponds to the dissipative regime). In this system, a hybrid regime emerges, with the Seebeck coefficient on one edge following the Kelvin formula and on the opposite edge following the Mott formula. The hybrid ballistic-dissipative thermoelectricity manifests itself in the anomalous Nernst effect. Finally, we examine the case of the Anderson topological insulator, showing that a non-magnetic disorder can induce a sharp change in the temperature dependence of the thermopower in the strong localization regime.  

The paper is organized as follows. In section II, we investigate the thermoelectric coefficients in a two-dimensional anomalous Hall system, specifically a Chern insulator. In section III, we study a two-dimensional topological insulator, considering both the cases of absence and presence of inelastic scattering of edge modes. Section IV focuses on the thermopower of a two-dimensional Anderson topological insulator. Finally, section V summarizes the main findings, discusses the potential for experimental verification of the results and their possible extension to other topological systems. Throughout the paper, we set $k_B=1$.

\section{Thermopower in an anomalous Hall system}

\begin{figure}[t]
\centering
\includegraphics[width = 0.8 \linewidth]{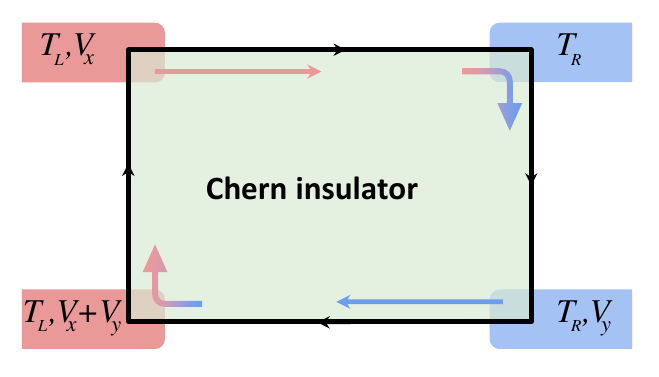}
\caption{Schematic representation of the four-terminal experimental setup. The reservoirs are connected to a two-dimensional Chern insulator of rectangular form. The closed edge mode is depicted with black color, while the red and blue arrows presents the termoelectric flow due to temperature and voltage differences in reservoirs.}
\label{Chern setup}
\end{figure}

We start by considering a two-dimensional system with anomalous Hall effect (AHE). A special case of such a system is the so-called two-dimensional Chern insulator (it is known as a class D in the topological classification~\cite{chiu2016classification}), which is an insulator in the bulk and has one gapless chiral edge channel (see Fig.~\ref{Chern setup}). Our goal is to study the thermopower of such a system and to verify the correctness of known thermoelectric relations such as the Kelvin, Mott, etc. formulas. The results obtained for such a system will then be generalized to the case of a two-dimensional topological insulator.

In general, it is well known that the electric current in a one-dimensional channel between right ($L$) and left ($R$) contacts 
in two-terminal geometry is given by the Landauer formula. The current is defined by the difference of two fluxes~\cite{landauer1970electrical,fisher1981relation}
\begin{equation}
I=\frac{2e}{h}\int d\epsilon\left[\Gamma_{LR}\left(\epsilon\right)f_L(\epsilon)-\Gamma_{RL}(\epsilon) f_R(\epsilon)\right], \label{Current_general}
\end{equation}
where $\Gamma_{LR}\left(\epsilon\right)$/$ \Gamma_{RL}\left(\epsilon\right)$ is the  barrier transparency from $L$/$R$ to $R$/$L$, and $f_{L/R}(\epsilon)=1/[1+\exp(\beta_{L/R}(\epsilon-\mu_{L/R}))]$ is is the Fermi-Dirac distribution function, $\mu_{L/R}$ and $T_{L/R}$ is the chemical potential and temperature of corresponding reservoir, and $\beta_{L/R}=1/T_{L/R}$ ia an inverse temperature. For a symmetric ballistic channel one has $\Gamma_{LR}=\Gamma_{RL}=1$.  

In a Chern insulator, the edge modes are ballistic, thus there is no scattering, which is due to the violation of T-symmetry and topological protection by Chern numbers. In contrast, in the case of a topological insulator, two scenarios are possible: the ballistic one, where the edge length is shorter than the inelastic scattering length, and the dissipative one that is realized if the edge length is longer than the inelastic scattering length. 

We will consider this in detail below. In the case of a Chern insulator, for the upper chiral channel from Fig.~\ref{Chern setup} we obtain $\Gamma_{LR}=1,\,\,\Gamma_{RL}=0$, and for the lower channel - $\Gamma_{LR}=0,\,\,\Gamma_{RL}=1$. Consequently, for the currents flowing through the upper ($u$) and lower ($l$) channels we find
\begin{gather}
I_{u/l}=\pm \frac{2e}{h}\int d\epsilon f_{L/R}(\epsilon).
\end{gather}
It is important to note that the currents created by the left and right reservoirs are separated in space. Since the reservoirs temperatures are different, we obtain spatially separated "hot" and "cold" currents. The realization of such kind of permanent ballistic electron transport has been proposed in the recent work by one of the present authors~\cite{Kavokin2}. 

When chiral edge modes are present, the so-called anomalous Nernst effect arises, which is similar to the conventional Nernst effect but occurs in the absence of external magnetic field. In this scenario, the Berry curvature replaces the magnetic field, generating what is nowadays known as the anomalous velocity. This phenomenon results in an additional carrier transfer between the edges and induces an extra potential difference, denoted further as $\Delta V_{NE}$. This potential difference must be accounted for, in addition to the thermopower.

Let us consider a two-dimensional electron system characterised by the energy spectrum $\varepsilon\left(p\right)$ and the Berry curvature $\bm{\mathcal{B}}_p$. The Berry curvature is a function of momentum and it is defined by the Bloch functions $u_{\bm{p}}$, i.e. $\bm{\mathcal{B}}_p=i\nabla_{\bm{p}}\times\bra{u_{\bm{p}}}\nabla_{\bm{p}}\ket{u_{\bm{p}}}$. The non-zero Berry curvature results in the semiclassical equations of motion having the form~\cite{xiao2010berry,Stephanov} 
\begin{gather}
\frac{d\bm{r}}{dt}=\bm{\upsilon}-\frac{1}{\hbar}\frac{d\bm{p}}{dt}\times\bm{\mathcal{B}}_p,\label{motion eq 1}\\
\frac{d\bm{p}}{dt}=e\bm{E}-e\nabla U \label{motion eq 2},
\end{gather}
where $\bm{\upsilon}=\partial\varepsilon/\partial\bm{p}$ is the group velocity of the bulk modes. The sign of the Berry curvature $\mathcal{B}_p$ in $\mathbb{Z}_2$ insulators is determined by the projection of the spin onto the momentum (chirality). The total Berry curvature for such a system, summed over both chiralities, vanishes, which is a consequence of time-reversal symmetry (T-symmetry). In contrast, in the Chern insulator under consideration, the Berry curvature takes a single value because of the violation of T-symmetry (quantum anomalous Hall effect).  In Eq.~\eqref{motion eq 2}, 
$\bm{E}$ represents an external electric field. Additionally, in a finite system, there is an edge lateral field, $U$. This field helps keeping the electrons inside the system. The gradient of it, $-\nabla U$, is directed perpendicular to the edge and it is nonzero only at the edge. By including the contribution of this field in the equation of motion above, we obtain
\begin{equation}
\frac{d\bm{r}}{dt}=\bm{\upsilon}+\bm{\upsilon}_{dr}-\frac{e}{\hbar}\bm{E}\times\bm{\mathcal{B}}_p,  
\end{equation}
where $\bm{\upsilon}_{dr}=-(e/\hbar)\left[\bm{\mathbf{\nabla}}U\times\bm{\mathcal{B}}_p\right]$. The second term on the right-hand side represents the effective velocity of the edge mode in the Chern insulator. The third term accounts for the anomalous Hall effect, which describes the generation of a transverse current induced by an external electric field applied in the absence of external magnetic field. This term is responsible for the emergence of the anomalous Nernst effect. For insulators, where the chemical potential belongs to the energy gap, we obtain $\bm{\upsilon}=0$ (localized states in the bulk). We shall assume this condition to be valid in the remaining part of this section.

The Berry curvature of a two-dimensional Chern insulator is oriented normally to the $XY$ plane, i.e $\bm{\mathcal{B}}_p=\bm{e}_{z}\mathcal{B}_p$, where $\bm{e}_{z}$ is the unit vector along axis $Z$. Therefore, $x-$components of the velocities of flows propagating along the opposite edges of the sample in the absence of additional external fields are given by 
\begin{equation}
\frac{d}{dt} x_{edge}^{u/l}= \pm \frac{e}{\hbar}\left|\nabla U\right|\mathcal{B}_p.
\end{equation}
The similar equations apply for $y$-component of velocities if the broken circuit condition is verified in all directions. 

Since the system is non-dissipative, the total current is given by the Landauer formula. The longitudinal current $I_{x}$ in this case includes two terms, which correspond to current flows originated from L and R reservoirs characterized by different chemical potentials $\mu_{L}$ and $\mu_R$ in general. For example, in the presence of the anomalous Hall effect, the chemical potentials for the upper and lower reservoirs may differ. Thus
\begin{gather}
I_{x}=I_{L}+I_{R},
\label{total current1}
\end{gather}
The contributions to the current on the right hand side of this expression are given by the following integrals
\begin{gather}
I_{L/R}=e\int_{-\pi}^{\pi}\frac{dk_{y}}{2\pi}\int\frac{dp_{x}}{2\pi\hbar}\left( \pm \upsilon_{dr}-\frac{e}{\hbar}E_{y}\mathcal{B}_p\right)f_{L/R}.
\label{Currents from left and right reservoirs1}
\end{gather}
Now, substituting Eq.~(\ref{Currents from left and right reservoirs1}) into Eq.~(\ref{total current1}) we arrive to the integral expression for the total current 
\begin{gather}
I_x = e \int\frac{d^{2}\bm{p}}{\left(2\pi\hbar\right)^{2}} \left( \upsilon_{dr} F_{-}-\frac{e}{\hbar}E_y \mathcal{B}_{p} F_{+}\right),
\label{total current}
\end{gather}
where $F_{\mp}= f_L \mp f_R$. Thus, the first term, associated with $F_{-}$ represents a pure Landauer contribution, while the second term corresponding to $F_{+}$, arises from the effective magnetic field appearing due to the Berry curvature, which is coupled to the total left (L) and right (R) particle densities. It is important to note that the second contribution vanishes in the absence of the transverse electric field component, $E_y$. In the following we shall aim at the induction of the transverse component of the field in order to reveal the impact of the Berry curvature on the termoelectric quantities. This can be achieved in a four-terminal setup schematically shown in Fig.~\ref{Chern setup}, as discussed below.   

First, we consider the general case of a system with an arbitrary non-zero Berry curvature. We then focus on a specific model. The Berry curvature significantly influences the carrier dynamics if the Fermi level is situated within the band gap, i.e., $-\Delta < \mu < \Delta$, where $2\Delta$ is the bandgap width. This occurs because the conduction and valence bands contribute to the total curvature with opposite-sign terms, and if $\mu> \Delta$, the contributions from both bands cancel each other. Conversely, for $\mu < - \Delta$, the Berry curvature is determined by the empty states, for which the distribution function vanishes. Let us introduce the following notations for temperatures of the left and right reservoirs $T_{L}=T+\Delta T$ and $T_{R}=T$ and consider the response regime that is linear in temperature variation, i.e. 
\begin{gather}
\label{chem poten}\mu_{L}=\mu+\frac{\partial\mu}{\partial T}\Delta T+eV_{x},\,\mu_{R}=\mu+eV_{y},
\end{gather}
where $V_{x}=E_{x}L_{x}$ and $V_{y}=E_{y}L_{y}$. In the above equation, the chemical potential $\mu_{L/R}$ corresponds to the upper left and lower right reservoirs. These reservoirs generate the thermoelectric flows depicted in Fig.~\ref{Chern setup} by the red and blue arrows. Relations~\eqref{chem poten} are characteristic of the considered regime where the "hot" and "cold" flows are spatially separated. In this regime the (L) and (R) chemical potentials are corrected with induced voltages  $V_x$ and $V_y$, respectively. For the convenience of further consideration, we set the potential of the upper right reservoir to zero. The potentials $V_{x/y}$ are then defined with respect to this reference value. For clarity, we ascribe the potential of $V_x$to the upper left reservoir, the potential of $V_x+V_y$ to the lower left reservoir, and the potential of $V_y$ to the lower right reservoir as shown in Fig.\ref{Chern setup} for a four-terminal setup. Now substituting Eq.~(\ref{chem poten}) into Eq.~(\ref{total current})  after straightforward algebra in the linear approximation we arrive to the following compact results
\begin{equation}
I_x = \sigma V_x - (\sigma+\sigma_H) V_y + \left(\alpha+\sigma\frac{1}{e}\frac{\partial \mu}{\partial T}\right) \Delta T.
\end{equation}
The thermoelectric coefficients in the above equation are given by the following integrals  
\begin{gather}
\sigma=e^2\int\frac{d^{2}p}{\left(2\pi\hbar\right)^{2}}\upsilon_{dr}\frac{\partial f_0}{\partial\mu},\label{longitudinal}\\
\sigma_{H}=\frac{2e^2}{\hbar}\int\frac{d^{2}p}{\left(2\pi\hbar\right)^{2}}\mathcal{B}_{p}f_0,\label{Hall plato}\\
\alpha=\frac{e}{h}\int\frac{d^{2}p}{\left(2\pi\hbar\right)^{2}}\upsilon_{dr}\frac{\varepsilon-\mu}{T}\left(-\frac{\partial f_{0}}{\partial\varepsilon}\right),\label{alpha}
\end{gather}
where $f_0=1/[1+\exp((\epsilon_p-\mu)/T)]$ is the equilibrium Fermi-Dirac distribution function at $V_x=V_y=0$ and $\Delta T=0$. Along the similar lines one can calculate the transverse current 
\begin{equation}
I_y = (\sigma+\sigma_H) V_x + \sigma V_y + \left(\alpha_{yx}+\sigma\frac{1}{e} \frac{\partial \mu}{\partial T} \right) \Delta T.
\end{equation}

The thermoelectric coefficient $\alpha_{yx}$, which is asymmetric with respect to the exchange $x\rightleftharpoons y$ is proportional to $\partial\sigma_{yx}/\partial\mu$. Our primary focus will be on the scenario where the chemical potential lies within the band gap, and the Hall conductivity exhibits only a weak dependence on the chemical potential. As a result, this coefficient is nearly zero, and we can safely neglect it. Solving the system of equations with $I_{x}=0$ and $I_{y}=0$, we obtain 
\begin{gather}
S=\frac{V_{x}}{\Delta T}=S_0+\delta S,\label{Seebeck}\\
S_0=-\frac{\sigma(2\sigma+\sigma_{H})}{\sigma^2+(\sigma+\sigma_H)^2}\frac{1}{e}\frac{\partial\mu}{\partial T},\label{Seebeck0}\\
\delta S=-\frac{\sigma\alpha}{\sigma^{2}+\left(\sigma+\sigma_{H}\right)^{2}}.
\label{Anomalous Seebeck}
\end{gather}
The Nernst coefficient is given by 
\begin{gather}
N=\frac{V_{y}}{\Delta T}=N_0+\delta N,\\
N_0=\frac{\sigma \sigma_H}{\sigma^2+(\sigma+\sigma_H)^2}\frac{1}{e}\frac{\partial\mu}{\partial T},\label{Nernst_Chern}\\
\delta N=-\frac{\sigma+\sigma_{H}}{\sigma}\delta S=\frac{\alpha\left(\sigma+\sigma_{H}\right)}{\sigma^{2}+\left(\sigma+\sigma_{H}\right)^{2}}.
\end{gather}
Next, we will obtain the approximate analytical expressions for $\delta S$ and $\delta N$. If we neglect the dependence of the Berry curvature on energy, it is easy to show that $\alpha=0$. In the two-terminal setup with only left and right electrodes remaining, it is necessary to set $V_y=0$. Than $S=-(1/e) \partial\mu/\partial T$ and $N=0$. One can conclude that in this case the thermopower is determined exactly by the Kelvin formula, where $\mu\left(T\right)$ is the chemical potential of the reservoir as a function of temperature. Note that for the two-terminal and four-terminal cases the thermopower is dependent of different factors, and that the Nernst coefficient in the first case vanishes. Note also that in the case of a Chern insulator, where the Hall conductivity is quantized, we obtain $\sigma=0$. In fact, in this case $\sigma_H=2e^2/h$, and $\sigma\propto\partial\sigma_H/\partial\mu=0$. Therefore, only one of the given thermoelectric coefficients does not vanish, namely $\delta N=\alpha/\sigma_H$. It is important to emphasize that the dependence of $N$ on the chemical potential is determined by the Berry curvature. Therefore, measurements of the Nernst coefficient can be proposed as an additional tool to study this important characteristic of Chern insulators. However, one should keep in mind that this result is valid, strictly speaking, at low temperatures and for large values of the band gap, when the Hall plateau exists. 

\begin{figure}[t]
\centering
\includegraphics[width = 0.8 \linewidth]{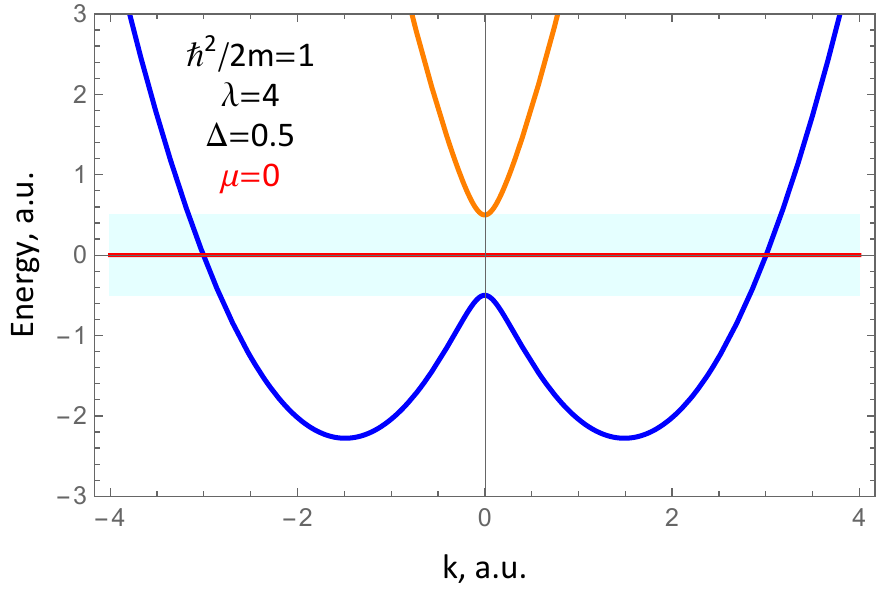}
\includegraphics[width = 0.85 \linewidth]{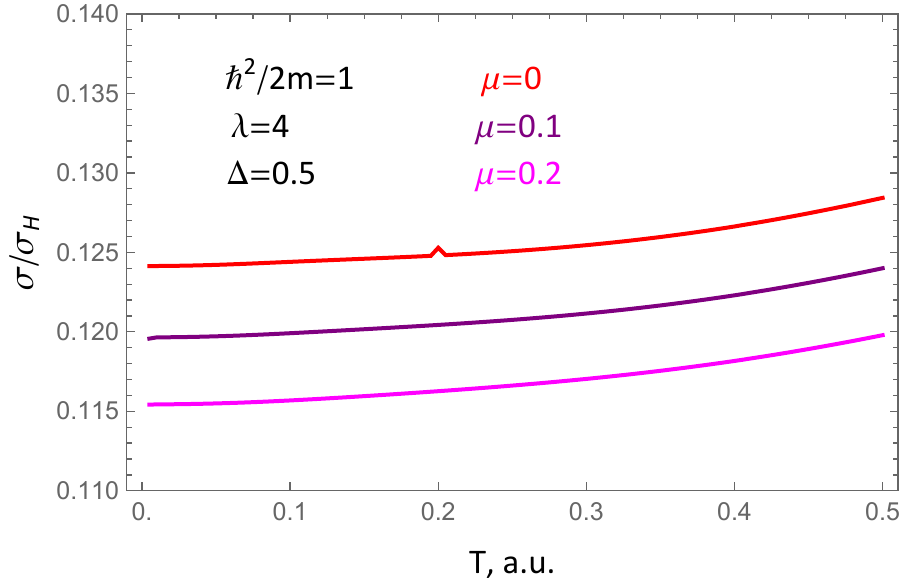}
\caption{Top: the energy spectrum of the Hamiltonian~\eqref{Rashba}. The region between the bands is highlighted with cyan light. Bottom: the dependence of $\sigma/\sigma_H$ on temperature. Temperature is given in the same units as $\Delta$.}\label{AHE and sigma}
\end{figure}

Until now we have developed a general formalism for a system that exhibits an anomalous Hall effect (the Chern insulator is a special case of such a system). The results obtained above are a consequence of only one assumption: the edge modes are ballistic. Now, we generalize this consideration for a specific model system. Namely, we shall consider a two-dimensional electron gas characterized with a Rashba spin-orbit interaction~\cite{onoda2006intrinsic}, given by the well-known Hamiltonian
\begin{equation}
H=\frac{\hbar^{2}k^2}{2m}+\lambda\left(\bm{k}\times\bm{\sigma}\right)\cdot\bm{e}_{z}-\Delta \sigma_{z}\label{Rashba},
\end{equation}
where $\lambda$ is the spin-orbit coupling constant, $2\Delta$ is the spin-orbit splitting of bands with different spin projections, $\sigma_z$ is the Pauli matrix. In this case, the Berry curvature is related to the spin-orbit interaction. Since the chemical potential is localized inside the gap, only the valence band gives a non-zero contribution to the Berry curvature. Inside the band gap it has the form 
\begin{equation}
\mathcal{B}_k=\frac{\lambda^{2}\Delta}{2\left(\lambda^{2}k^2+\Delta^{2}\right)^{3/2}}.\label{Berry}
\end{equation}
Given that the energy spectrum in this case is $\varepsilon_{\pm}=\hbar^{2}k^{2}/2m\pm\sqrt{\lambda^{2}k^{2}+\Delta^{2}}$, we obtain $\mathcal{B}_k=\mathcal{B}_{k\left(\varepsilon_{-}\right)}$. This leads to the following expression
\begin{gather}
\alpha=\frac{e^2}{\hbar}\left|\nabla U\right|\int_{-\pi}^{\pi}\frac{dk_{y}}{2\pi}\int\frac{dk_{x}}{2\pi}\mathcal{B}_{k\left(\varepsilon\right)}\frac{\varepsilon-\mu}{T}\frac{\partial f_{0}}{\partial\mu}.
\end{gather}
In the above expression and further on, we omit the subscript $''-''$ for energy. Using the expression for the Berry curvature $\mathcal{B}\left(\varepsilon\right)$ one can obtain an analytical expression for the anomalous part of the thermopower. At low temperatures ($\mu \gg T$):
\begin{gather}
\delta S=\frac{1}{e}\frac{\pi^2 T}{3}\frac{\partial \ln\left(\rho\mathcal{B}\right)_{\varepsilon=\mu}}{\partial\mu} \frac{1}{1+\left(1+\sigma_{H}/\sigma\right)^{2}},
\end{gather}
where
\begin{gather}
\rho\left(\varepsilon\right)=\frac{2m}{\hbar^{2}}\left[1+\frac{\lambda^{2}}{\sqrt{2\frac{\hbar^{2}\varepsilon}{m}\lambda^{2}+\lambda^{4}+\left(\frac{\hbar^{2}}{m}\right)^{2}\Delta^{2}}}\right].\label{DoS Rashba}
\end{gather}
The Hamiltonian, Eq.~\eqref{Rashba}, under consideration describes an anomalous but still classical Hall effect. Therefore, generally speaking, both coefficients vanish. To study them, we note once again that in topological systems an important role is played by the ratio $\sigma/\sigma_H$, which enters the expressions for the thermoelectric coefficients. Figure~\ref{AHE and sigma} shows the temperature dependence of this quantity calculated with use of expressions ~\eqref{longitudinal},~\eqref{Hall plato},~\eqref{Berry},~\eqref{DoS Rashba}). One can see that $\sigma/\sigma_H$ only slightly depends on temperature over a wide temperature range. Consequently, at low temperatures the Seebeck and Nernst coefficients are linear functions temperature, in a good agreement with the Mott formula.

In Ref.~\cite{Kavokin2} it was shown that an anomalous correction to the Kelvin formula appears in graphene. This anomaly is related to the fact that the quantity $\upsilon\rho$ is a function of energy due to the linear electronic dispersion close to the Dirac point. In our case, the anomalous velocity is governed by the Berry curvature. Therefore, the anomaly in the thermopower will be determined by the quantity $\rho\mathcal{B}$.

Finally, we point out that the AHE regime can be realized in a 3D topological insulator with ferromagnetic plates on the upper and lower sides (see, for example,~\cite{Qi}). In our opinion, such a system is the most convenient one for thermoelectric measurements. In the next section, we will consider the thermoelectric properties of a 2D topological insulator without T-symmetry breaking.

\section{Thermopower of 2D topological insulator}

\begin{figure}[t]
\includegraphics[width = 0.7 \linewidth]{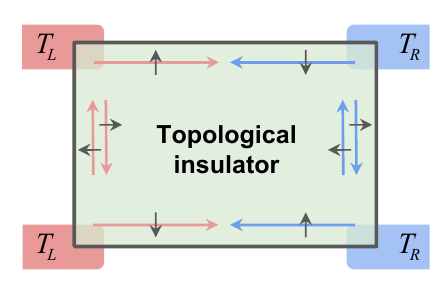}
\caption{The schematic of the system under study: a two-dimensional topological insulator is sandwiched between reservoirs of different temperatures. The arrows along the edges show the directions of the two chiral modes, and the small arrows show the spin projections.}
\label{fig_TI}
\end{figure}

In this section, we generalize the results obtained above to the case of a two-dimensional topological insulator (class AII in the topological classification~\cite{chiu2016classification}). The simplest effective Hamiltonian describing the edge states has the form: $H_{eff}=\int\sum_{s=\pm}\frac{dk_x}{2\pi} s \upsilon k_{x}\psi^{\dagger}_{k_{x}s}\psi_{k_{x}s}$. Fig.~\ref{fig_TI} shows a two-dimensional TI with four reservoirs (four-terminal scheme). At the edges, there are one-dimensional spin-degenerate chiral modes: the upper channel is characterised by positive chirality (spin up carriers propagate to the right, and spin down carriers propagate to the left), and the lower channel is characterised by negative chirality (spin up carriers propagate to the left, and spin down ones propagate to the right). We assume that the chemical potential is located inside the band gap, which is considered wide enough to neglect bulk states.

We first consider the case where the edge modes are ballistic (in the next section, the effects of inelastic scattering will be taken into account). Thus, the current in the upper channel is written as
\begin{gather}
I_{u}=\frac{2e}{\hbar}\int d\epsilon\left[f_{\uparrow}\left(\mu_{L},T_{L}\right)-f_{\downarrow}\left(\mu_{R},T_{R}\right)\right],
\end{gather}
and in the lower one, respectively
\begin{gather}
I_{l}=\frac{2e}{\hbar}\int d\epsilon\left[f_{\downarrow}\left(\mu_{L},T_{L}\right)-f_{\uparrow}\left(\mu_{R},T_{R}\right)\right].
\end{gather}

In terms of the anomalous velocity, in the case under consideration, the sign of the Berry curvature is determined by the spin projection: $\mathcal{B}=\text{sign}\left(s\right)\left|\mathcal{B}\right|$.
In this regime, T-symmetry is not broken and there is no spin splitting. This is why, $f_{\uparrow}=f_{\downarrow}$ and both channels are absolutely equivalent. Therefore, using the same approach as in the previous section, we obtain $V_{x}=-\frac{1}{e}\frac{\partial \mu}{\partial T}\Delta T$ and $V_{y}=0$. In this case the thermopower is given by the Kelvin formula. This is clear from symmetry considerations. One can also prove this statement by a direct calculation of the current using Eqs~\eqref{chem poten}. To recover the previous case of chiral spin-nondegenerate modes (Chern insulator), it is necessary to induce the spin splitting of carriers. This can be done by introduction of a magnetic field, magnetic impurities, or by using completely spin-polarized electrodes (ferromagnetic reservoirs). The latter case is realized in half-metals, where the density of states for one spin projection vanishes completely at the Fermi level. The mechanisms of generation of pure spin currents are discussed, for example, in Refs~\cite{Takahashi, Adachi, Valenzuela, Friso}. In the following, we discuss the regime where chiral edge modes in a topological insulator become dissipative.

\subsection{Inelasting scattering of edge modes}

Topological insulators are characterised by chiral channels protected from scattering by T-symmetry. These modes sustain the ballistic transport of electrons. The corresponding conductivities are determined by fundamental constants, as in any one-dimensional ballistic channel. However, experimentally the conductivities of edge modes frequently deviate from the analytical predictions ~\cite{konig2007quantum,konig2013spatially,olshanetsky2023observation}. Such deviations are due to the presence of inelastic scattering of edge modes. A number of mechanisms for such scattering has been proposed. In particular, the inelastic scattering is possible in systems with bulk inversion asymmetry (BIA) and structural inversion asymmetry (SIA)~\cite{schmidt2012inelastic}. Inclusion of such types of asymmetry on one hand does not violate T-symmetry (and therefore, the gap in the energy spectrum does not open), but on the other hand it leads to the mixing of channels characterised by different spin projections, so that counter-propagating edge modes no longer possess well-defined spin polarizations~\cite{rothe2010fingerprint, maciejko2010magnetoconductance}. This inversion symmetry breaking does not impact on the chiral transport in the case of elastic scattering of carriers, since such transport is protected by T-symmetry. However, in the case of inelastic scattering this can lead to a non-zero overlap between states with different momenta. This in turn affects the conductivity of the edge state~\cite{schmidt2012inelastic}. In Ref~\cite{du2015robust} this was demonstrated experimentally. In addition, a mechanism of inelastic scattering of edge modes by charge puddles formed in a two-dimensional TI upon doping was proposed~\cite{vayrynen2013helical,vayrynen2014resistance}. In the present study, we will capitalize on the results of Ref~\cite{schmidt2012inelastic}. It demonstrated that for a two-dimensional topological insulator based on an InAs/GaSb quantum well, the resistance of edge modes increases linearly with increasing sample length $l$, starting from $l_\phi=4.4 \mu m$ ($l_{\phi}$ is the inelastic scattering length). For $l<l_\phi$, the depenence of the resistance on the sample length exhibits a plateau, which certifies that the edge transport is ballistic. Hence, for $l>l_\phi$, the edge mode undergoes inelastic scattering. Below, we present calculations based on the microscopic theory of inelastic scattering of edge modes.

\begin{figure}[t]
\centering
\includegraphics[width = \linewidth]{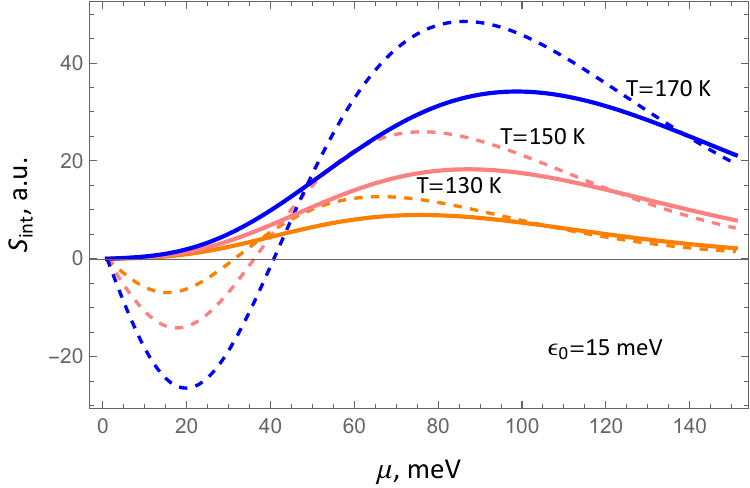}
\caption{The Seebeck coefficient~\eqref{Sommerfeld} plotted as a function of the chemical potential at different temperatures (solid lines). $S_{int}\left(\frac{\pi^2}{3e}L k_0\left(\frac{U_0}{\upsilon_F}\right)^2 \right)^{-1}$. For comparison, the curves obtained from the formula~\eqref{Seebeck inelastic} (dashed lines) are also shown. As one can see, at a certain value of $\mu/T$ the thermopower changes sign, according to the Mott formula.}
\label{Inelastic}
\end{figure}

Let us consider the case where the longitudinal size of the system is larger than the inelastic scattering length, and the transverse size is smaller. Thus, the longitudinal edge modes (at the upper and lower edges, see Fig.~\ref{fig_TI}) experience scattering, and the transverse ones experience ballistic modes. To calculate the Seebeck coefficient in the presence of scattering, we will use the standard expression (in literature this formula is sometimes called the Sommerfeld-Bethe relation~\cite{sommerfeld2013elektronentheorie,mott1958theory,wilson2011theory})
\begin{gather}
S=-\frac{1}{eT}\frac{\int_{-\infty}^{\infty}\left(\varepsilon-\mu\right)\left(-\frac{\partial f_{0}}{\partial\varepsilon}\right)\sigma\left(\varepsilon,T\right)d\varepsilon}{\int_{-\infty}^{\infty}\left(-\frac{\partial f_{0}}{\partial\varepsilon}\right)\sigma\left(\varepsilon,T\right)d\varepsilon}\label{Sommerfeld}.
\end{gather}
At low temperatures we recover the Mott formula
\begin{gather}
S\approx-\frac{\pi^2 T}{3e}\frac{\partial\ln\sigma\left(\mu,T\right)}{\partial\mu}.\label{Mott inelastic}
\end{gather}
Next, we will use the results of Ref~\cite{schmidt2012inelastic} for the conductivity in the presence of inelastic scattering
\begin{gather}
\sigma\left(\mu,T\right)\approx\frac{e^2}{h}+\delta\sigma\left(\mu,T\right),
\end{gather}
where $\delta\sigma$ is given by
\begin{gather}
\delta\sigma\left(\mu,T\right)=\frac{e^2}{h}L k_0\left(\frac{U_0}{\upsilon_F}\right)^2 \left(\frac{T}{\epsilon_0}\right)^5 \mathcal{F}\left(\frac{\mu}{T}\right),
\end{gather}
where
\begin{gather}
\mathcal{F}\left(x\right)=\frac{8}{\pi}\int_{-\infty}^{\infty}dx_1 dx_2 \left(x_1^2 - x_2^2\right)^2 f_0\left(x_1 - x\right)f_0\left(x_2 - x\right)\nonumber\\
\times\left[1-f_0\left(x_1+x_2 - x\right)\right]\left[1-f_0\left(-x\right)\right],
\end{gather}
where $\epsilon_0=\upsilon_F k_0$, and other symbols are the same as in Ref.~\cite{schmidt2012inelastic}. Substituting these expressions into Eq~\eqref{Mott inelastic}, we obtain an expression for the Seebeck coefficient. Taking into account that $\delta\sigma<e^2/h$, we can neglect this correction in the denominator of Eq.~\eqref{Mott inelastic}. Then
\begin{gather}
S_{in}=-\frac{\pi^2}{3e}L k_0\left(\frac{U_0}{\upsilon_F}\right)^2 \left(\frac{T}{\epsilon_0}\right)^5 \mathcal{F}'\left(\frac{\mu}{T}\right).\label{Seebeck inelastic}
\end{gather}
Fig.~\ref{Inelastic} shows the function $S_{in}\left(\mu\right)$ plotted with use of Eq.~\eqref{Sommerfeld}. For the sake of comparison, we also show the predictions of the Mott formula. One can see that the Mott formula predicts the change of the sign of the thermopower at a certain value of $\mu/T$. However, this prediction is misleading, since it the change of the sign is expected in the region where the low-temperature approximation is no longer valid. 

\begin{figure}[t]
\includegraphics[width = \linewidth]{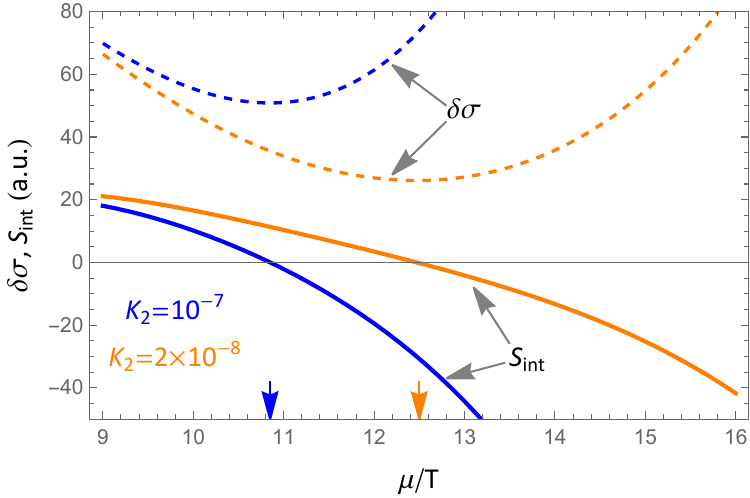}
\caption{Dependencies of $\delta\sigma$ (dashed lines) and $S_{int}$ (solid lines) as functions of $\mu/T$ for different values of the parameter $K_2$. At some value of $\mu/T$ corresponding to the low-temperature approximation, the Seebeck coefficient changes its sign.}
\label{Inelastic+}
\end{figure}

Further, we note that a different mechanism of inelastic scattering of edge modes was considered in Ref~\cite{schmidt2012inelastic}. This mechanism arises as a single-particle backscattering process in the second order in the electron-electron interaction and impurity potential. In the paper~\cite{schmidt2012inelastic}, a low-temperature expression for the correction to the conductivity in the presence of such scattering was obtained. This correction has nontrivial consequences in what concerns the thermopower calculation. Indeed, in the presence of the two above mentioned mechanisms, the Zebeeck coefficient changes sign at a certain value of the Fermi level at low impurity concentrations. This follows from the Mott formula, which is correct in this case, since the effect arises in the low-temperature region. Taking into account both mechanisms, we write the correction to the conductivity using Eqs. (11, 12) from~\cite{schmidt2012inelastic}):
\begin{gather}
\delta\sigma=\frac{e^2}{h}K_1 \left(\frac{\mu}{T}\right)^6 \left(e^{-\frac{\mu}{T}}+K_2 \left(\frac{\mu}{T}\right)^2 \right),
\end{gather}
where $K_1\approx\frac{1}{\pi} L k_0\left(\frac{U_0}{\upsilon_F}\right)^2 \left(\frac{T}{\upsilon_F k_0}\right)^5$, $K_2\approx10^4 \frac{n_{imp}}{k_0} \left(\frac{V_0}{\upsilon_F}\right)^2 \left(\frac{T}{\upsilon_F k_0}\right)^7$. Using this expression we can find the Seebeck coefficient from the Mott formula~\eqref{Mott inelastic}. The corresponding curves are shown in Fig.~\ref{Inelastic+}. The figure shows the correction to the conductivity, as well as the Seebeck coefficient. One can see that at a certain value of $\mu/T$ (at which the low-temperature approximation can still be considered good enough!) the Seebeck coefficient changes sign.

\subsection{Hybrid thermoelectric regime in the trapezoid geometry}

\begin{figure}[t]
\centering
\includegraphics[width = 0.7 \linewidth]{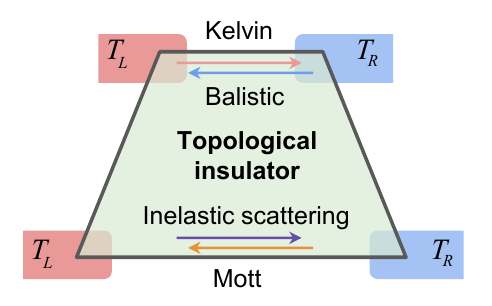}

\caption{Schematic of a trapezoidal system where different edges are characterized by different mode dissipation. We are interested in the situation where the short edge sustains ballistic modes, while the long edge can only sustain modes that undergo inelastic scattering.}
\label{Trapezoid}
\end{figure}

Let us now consider the system shown in Fig.~\ref{Trapezoid}. This is a two-dimensional topological insulator in the form of a trapezoid. The relationship between the upper and lower bases is such that the upper edge mode is ballistic ($L<l_\phi$), and the lower edge mode is subject to scattering ($L>l_\phi$). Then at the upper edge the Seebeck coefficient is given by the Kelvin formula, and at the lower edge it is given by the Mott formula~\eqref{Seebeck inelastic}. Thus, these opposite edges will be characterized by different thermopowers, which can lead to a non-zero Nernst coefficient. 

Following the first section, it is easy to show that the Seebeck coefficients at the two edges are generally defined as
\begin{gather}
S_u=-\frac{1}{e}\frac{\partial\mu_{u}}{\partial T}-\frac{\alpha_u}{\sigma_u},\\
S_l=-\frac{1}{e}\frac{\partial\mu_{l}}{\partial T}-\frac{\alpha_l}{\sigma_l},
\end{gather}
where indices $u,l$ denote the upper and lower edges. In the most interesting case, when the ballistic regime is realized at one (upper) edge, and inelastic scattering occurs at the second (lower) edge, we obtain: $S_{u}=-(1/e)\partial\mu/\partial T, S_l=-\alpha/\sigma$, where $\alpha,\sigma$ correspond to the lower edge. Let us underline that in the upper expression $\mu$ is the chemical potential of the reservoir. In the lower expression it is taken into account that the chemical potential of a one-dimensional channel does not depend on temperature. Thus, at one edge the thermopower is determined by the Kelvin formula, and at the other, at low temperatures, it is governed by the Mott formula. That is, at one edge it is determined by the temperature dependence of the chemical potential of the reservoirs, and at the second edge it depends on the system parameters. This means that electromotive forces of different magnitudes will be generated at different edges. This will lead to a non-zero Nernst voltage at the left and right edges of the system. It is easy to show that the voltages at different edges are of the same magnitude but opposite signs. The corresponding Nernst coefficients are of the form
\begin{gather}
N_L=\frac{S_l-S_u}{2}=\frac{1}{e}\frac{\partial\mu}{\partial T}-\frac{\alpha}{\sigma},\\
N_R=-N_L.
\end{gather}
In this case, the total transverse current vanishes, which corresponds to the open-loop condition of the system, as well as the presence of T-symmetry. Thus, the non-equivalence of the edges with respect to scattering leads to the appearance of a non-zero Nernst coefficient. Since inelastic scattering depends on the edge length, different values of the Nernst thermopower will be observed in different trapezoidal samples characterized by the same upper base but different lower bases. This effect offers an experimental tool for studying the dissipative properties of edge modes in a topological insulator.

In the same framework, one can consider more complex geometries where different scattering modes are also realized at the side edges (right and left).

\section{Thermopower of a 2D Anderson topological insulator}

One of the most topical issues in the field of topological systems is the effect of various types of disorder on topological phases~\cite{Liu2009,Groth,Guo,Meier}. Disorder plays a key role in establishing a number of nontrivial properties of electronic systems. For example, disorder is the cause of quantum localization (Anderson localization) and Hall plateaus in the integer quantum Hall effect~\cite{Tong}. In the previous sections, we have shown that disorder can be the cause of additional inelastic scattering of edge modes in topological insulators with SIA and BIA. Another interesting effect is disorder-induced topological phases in two- and three-dimensional systems, which have been termed two- and three-dimensional Anderson topological insulators. The disorder in these systems may induce the Anderson localization. On the other hand, due to this disorder, the topological mass changes it sign that is characteristic of a band inversion. As a consequence, the system transfers into a topologically nontrivial state~\cite{Groth}. Thus, the disorder induces chiral edge states in the system, which in the ideal state (in the pure limit) is topologically trivial. It should be noted that the topological Anderson phase was predicted not only for crystalline systems, but also for amorphous ones (see, e.g.,~\cite{XiaoyuCheng}).

Here we study the thermopower of a two-dimensional Anderson TI. It should be noted that a scaling theory of the thermoelectric response of two-dimensional systems with disorder has been recently developed~\cite{yamamoto2024scaling}. We consider the situation where a two-dimensional semiconductor system with spin-orbit interaction is initially in the trivial phase (positive band gap), and the chemical potential is located in the conduction band. The electronic transport is dominated by two-dimensional bulk modes, which are dissipative, in general. Introduction of a non-magnetic disorder into such a system leads to the localization of these modes, as well as to the renormalization of the band gap and chemical potential~\cite{Liu2009,Groth,Guo}. With increase in the defect concentration to a certain critical value, the band gap collapses and then reopens with the opposite sign. Then the system is characterized by a non-trivial topological phase with chiral edge modes (Anderson TI). Such a rearrangement leads to some anomaly in the thermopower, which we will now discuss in detail.

Following Refs~\cite{Liu2009,Groth,Guo,Meier}, we use the following 4-band Hamiltonian for a cubic lattice topological insulator in the presence with disorder:
\begin{equation}
H=\sum_{\mathbf{k}}\varPsi_{\mathbf{k}}^{\dagger}\left[d_{\mu}\left(\mathbf{k}\right)\Gamma_{\mu}+d_{4}\left(\mathbf{k}\right)\right]\varPsi_{\mathbf{k}}+\sum_{j}U_{j}\varPsi_{j}^{\dagger}\varPsi_{j},
\end{equation}
where $\mu=\left(0,1,2,3\right)$, $\varPsi_{j}$ is a 4-component state vector at the $j$th site of the crystal lattice (accordingly, $\varPsi_{\mathbf{k}}$ is the Fourier transform of this vector), $d_{0} \left(\mathbf{k}\right)=\chi-2t\sum_{i}\cos k_{i}$, $d_{i}\left(\mathbf{k}\right)=-2\lambda\sin k_{i}$, $d_{4}\left(\mathbf{k}\right)=2\gamma\sum_{i}\left(1-\cos k_{i}\right)$, $\chi,t,\lambda,\gamma$ are the tight-binding model parameters (see~\cite{Liu2009,Groth}), $\Gamma_{\mu}$ are the Dirac matrices, $U_{j}$ is the random potential at the $j$-th lattice site caused by disorder. For simplicity, the lattice constant is set equal to unity. We use the simplest Anderson model, in which the energy values at the sites are distributed uniformly with a density $1/U_{0}$ in the range of $\left[-U_{0}/2,U_{0}/2\right]$.

For $U_{0}=0$ we obtain the Hamiltonian of an ideal system $H_{TI}^{0}=\sum_{\mathbf{k}}\varPsi_{\mathbf{k}}^{\dagger}H_{\mathbf{k}}^{0}\varPsi_{\mathbf{k}}$. The term $m=\chi-6t$ at $\Gamma_0$ in this Hamiltonian is referred to as the topological mass. For $m>0$, the system is an ordinary band insulator with a bandgap $m$. If $m<0$, the spectrum becomes inverted, which leads to the appearance of additional states with zero energy, namely, the chiral edge modes. Note that the Hamiltonian is time-reversal symmetric since $\mathcal{T}\Gamma_{0}\mathcal{T}^{-1}=\Gamma_{0}, \mathcal{T}\Gamma_{i}\mathcal{T}^{-1}=-\Gamma_{i}, d_{i}\left(-\mathbf{k}\right)=-d_{i}\left(\mathbf{k}\right)$, where $\mathcal{T}$ is the time-reversal operator. This symmetry is responsible for the topological protection of edge modes. However, due to inelastic scattering this protection maybe suppressed in the presence of SIA and BIA, see the previous section.

The expression for the topological mass $m=\chi-6t$ is obtained in the zeroth order of the expansion of the function $d_{0}\left(\mathbf{k}\right)$ in a Taylor series near $\mathbf{k}=0$. In the second order, a quadratic term $tk^{2}$ appears, which plays an important role for the topological classification. In the presence of disorder ($U_{0}\neq0$) this term is even more significant. Indeed, disorder leads to spatial localization of states, where the extension of wave functions is governed by the localization length: $\psi\sim\exp\left(-r/r_{0}\right)$. The presence of the quadratic term leads to the correction to the topological mass: ($\hat{k}=-i\nabla$): $-t\nabla^{2}\psi=-tr_{0}^{-2}\psi$. This negative correction can become larger than the bandgap of the ideal system $m$. This leads to a change in the sign of the topological mass, i.e. an inversion of the spectrum. As a consequence, gapless states arise in the form of chiral edge modes. This effect plays a key role in establishing the so-called Anderson topological insulator phase~\cite{Guo,Liu2009,Groth}. 

The dependencies $\overline{m}\left(U_{0}\right)$ and $\overline{\mu}\left(U_{0}\right)$ worth a supplementary discussion. In an ideal system, if $U_0=0$, the trivial phase corresponds to $m>0$, and for $|\mu|<m$ this phase is a trivial band insulator. We consider the situation where $\mu>m$. Disorder leads to the renormalization of topological mass and chemical potential. The transition to the topological phase occurs through the gap closing. In the general case, we denote by $U_{c1}$ the value that $U_{0}$ takes where the topological mass vanishes, i.e. $\overline{m}\left(U_{c1}\right)=0$. Then $\overline{m}<0$ for $U_{0}>U_{c1}$. In addition, let us denote the value of $U_{0}$ by $U_{c2}$ where the renormalized chemical potential is located inside the band gap: $\overline{\mu}\left(U_0>U_{c2}\right)<\overline {m}$. Thus, one can select three important ranges of $U_0$: 1) $0<U_{0}<U_{c1}$, 2) $U_{c1}<U_{0}<U_{c2}$, 3) $U_{c2}<U_{0}$. 

In an infinite two-dimensional system, even a weak disorder leads to the localization of all states at zero temperature. At nonzero temperatures, the characteristic length scale is the dephasing length (diffusion length). At low temperatures, the extended states are still suppressed, the weak localization takes place. We will assume that our system is large enough to take advantage of the scaling theory of localization. Thus, at nonzero temperatures, the states of the conduction band (bulk states) of a two-dimensional system give a nonzero contribution to transport. Although the conductivity is close to zero, the contribution to the thermopower can be significant. Our goal is to study the thermopower in the coexistence of weakly localized bulk states and delocalized chiral edge modes. We will pay special attention to the transition temperature between the modes of weak and strong (Anderson) localization.

First, consider the region $0<U_{0}<U_{c1}$. In this region, the system under study contains only bulk modes, which are suppressed due to disorder. The most interesting is the transition from weak to strong localization~\cite{yamamoto2024scaling}. 

For the weak localization phase, the scaling theory tells us that the conductivity of the system is given by~\cite{lee1985disordered}
\begin{gather}
\sigma_{wl}\left(\varepsilon,T\right)=\sigma_{0}\left(\varepsilon\right)-\alpha\ln{\frac{L_{\phi}\left(\varepsilon,T\right)}{L}},
\end{gather}
where $\alpha$ is the expansion constant in the scaling theory~\cite{abrahams1979scaling}, $L_{\phi}$ is the so-called dephasing length due to inelastic scattering, $L$ is the linear size of the system (this quantity is considered to be larger than any mean free path), and $\sigma_0$ is the conductivity of a system of size $L$ at zero temperature (it is not zero because the system is finite in size). In the weak localization phase, $\sigma_0$ is not small. We write $L_{\phi}$ as $L_{\phi}\left(\varepsilon,T\right)=L_{0}\left(\varepsilon\right)\left(T/T_{0}\right)^{-p}$. In this expression we have separated the temperature part $L_{\phi}\left(\varepsilon,T\right)$, which is of key importance for this work. At the same time, the exact forms of the quantities $L_0$ and $T_0$ are not so important, since we are interested in a qualitative consideration of the temperature dependence of the Seebeck coefficient. Thus
\begin{gather}
\sigma_{wl}\left(\varepsilon,T\right)=\sigma_{0}\left(\varepsilon\right)-\alpha p\ln{\left(\frac{T_{wl}\left(\varepsilon\right)}{T}\right)},
\end{gather}
where we introduced the notation $T_{wl}=\left(L_0\left(\varepsilon\right)/L\right)^{1/p}T_0$, and the correction is negative for $T<T_{wl}$. The temperature of the transition to strong (Anderson) localization is estimated as $T_{wl-sl}\propto T_{wl}\exp\left(-\sigma_0/\alpha p\right)$. 

In the limit of strong localization for conductivity one can write (see, for example,~\cite{lee1985disordered})
\begin{gather}
\sigma_{sl}\left(\varepsilon,T\right)=\alpha\exp\left[-\left(\frac{T_{sl}\left(\varepsilon\right)}{T}\right)^{p}\right],
\end{gather}
where we have introduced the notation $T_{sl}=T_0 \left(\frac{L_0\left(\varepsilon\right)}{L}\ln\frac{\alpha}{\sigma_0}\right)^{1/p}$. One can see that the conductivity in the two phases (weak and strong localization) is significantly different. Therefore, the transition from one phase to another one must be accompanied by a resonant feature in the thermopower. For the Seebeck coefficient, we will use Eq.~\eqref{Sommerfeld}. We shall assume that the temperature is much lower than the chemical potential. Then, in the weak localization phase, we obtain
\begin{gather}
S_{wl}=-\frac{\pi^2}{3e}\frac{T}{T_{wl}}\frac{\sigma_{0}'\left(\mu\right)T_{wl}-\alpha p T_{wl}'}{\sigma_{0}\left(\mu\right)-\alpha p \ln{\left(\frac{T_{wl}}{T}\right)}}.
\end{gather}
If we neglect the weak correction to the conductivity in the denominator, we obtain
\begin{gather}
S_{wl}\approx-\frac{\pi^2 T}{3e} \left[\frac{\partial\ln{\sigma_{0}\left(\mu\right)}}{\partial\mu}-\frac{\alpha p}{\sigma_{0}\left(\mu\right)}\frac{\partial\ln{T_{sl}\left(\mu\right)}}{\partial\mu}\right].\label{weak localization}
\end{gather}
One can see that the thermopower is governed by the Mott formula corrected due to the effects of weak localization. In the strong localization limit we obtain
\begin{gather}
S_{sl}=\frac{\pi^2 p}{3e}T^{p}_{sl}\left(\mu\right) \frac{\partial\ln{T_{sl}\left(\mu\right)}}{\partial\mu} T^{1-p}.\label{strong localization trivial}
\end{gather}
Note that the temperature dependencies of the Seebeck coefficient differ in different phases. In addition, unlike conductivity the thermopower in the strong localization regime is not suppressed exponentially. 

\begin{figure}[t]
\includegraphics[width = \linewidth]{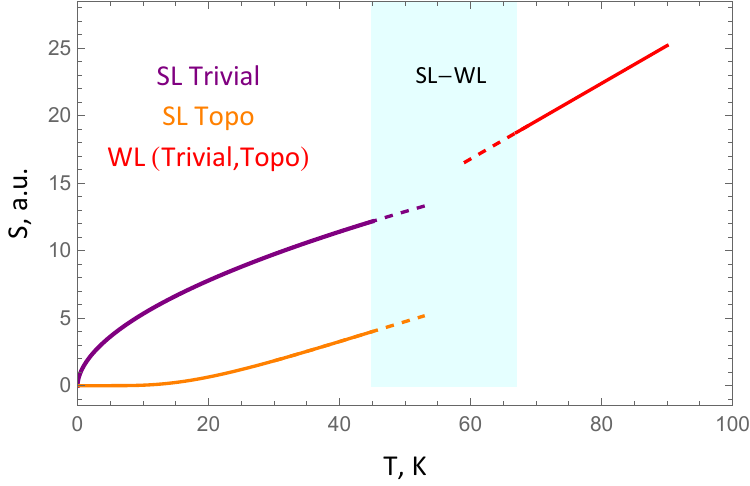}
\caption{The temperature dependence of thermopower in trivial ($0<U_{0}<U_{c1}$) and topological ($U_{c1}<U_{0}<U_{c2}$) phases to the left and right of the SL-WL transition. We use the following notations: SL Trivial - strong localization in the trivial phase (Eq.~\eqref{strong localization trivial}), SL Topo - strong localization in the topological phase (Eq.~\eqref{strong localization topo}), WL (Trivial,Topo) - weak localization in the trivial and topological phases (Eq.~\eqref{weak localization}). The SL-WL transition region is highlighted in cyan.}
\label{sl-wl}
\end{figure}

Now let us consider the region $U_{c1}<U_{0}<U_{c2}$. In this case, in addition to the nearly localized (strongly or weakly) bulk states, there is also a topological edge mode whose contribution to the current is determined using Eq.~\eqref{Current_general}. Note that the presence of the edge mode will have almost no effect on the thermopower in the weak localization regime. This is due to the fact that in this case the contribution from the bulk modes is dominant. However, in the strong localization regime, the appearance of the edge mode significantly affects the thermopower. Due to the exponential suppression of conductivity in this regime, the main contribution to the transport is given by the edge mode, and the bulk modes can be neglected. In this case, the conductivity is given by
\begin{gather}
\sigma_{sl}\left(\varepsilon,T\right)\approx\frac{e^2}{h}+\alpha\exp\left[-\left(\frac{T_{sl}\left(\varepsilon\right)}{T}\right)^{p}\right].
\end{gather}
Using the Mott's formula and neglecting the weak correction from localized bulk modes in the denominator, we obtain the following expression for the Seebeck coefficient:
\begin{gather}
S_{sl}\propto T^{1-p} \exp\left[-\left(\frac{T_{sl}\left(\varepsilon\right)}{T}\right)^{p}\right].\label{strong localization topo}
\end{gather}
One can see that in the presence of the edge mode, the thermopower is suppressed exponentially, similar to the conductivity. That is, the appearance of the edge mode leads to a dramatic change in the temperature dependence of the Seebeck coefficient: in the trivial phase $S_{sl}\propto T^{1-p}$, and in the topological $S_{sl}\propto T^{1-p} \exp\left(-\gamma T^{-p}\right) $. In the "weak localization" region, the temperature dependencies of the conductivity would not differ significantly in these two phases: $S_{wl}\propto T$. Thus, the disorder-induced transition to a topologically nontrivial phase leads to a significant change in the behavior of the thermopower near the weak-strong localization transition. Fig.~\ref{sl-wl} shows a qualitative picture of the temperature dependence of the thermo electromotive force in the trivial ($0<U_{0}<U_{c1}$) and topological ($U_{c1}<U_{0}<U_{c2}$) phases.

Finally, for $U_{0}>U_{c2}$ the bulk states are strongly suppressed because the chemical potential is located inside the gap. Thus, the bulk modes do not give any sizable contribution to the thermopower. In this regime, the thermopower is given by the expressions found in the previous sections.

\section{Conclusion}

In conclusion, we have examined the interplay of ballistic and diffusive electron transport in topological insulators. We analyzed the impact of their bulk and edge modes on the main thermoelectric and thermomagnetic coefficients: the thermopower or Seebeck coefficient and the Nernst coefficient. Our focus was on a specific type of non-magnetic scattering of edge modes. However, it is worth mentioning that there are other mechanisms, which can play an important role in thermoelectric effects~~\cite{dolcetto2016edge}) , and moreover recently, a different non-magnetic mechanism has been explored in Ref.~~\cite{krainov2024non}. Our results in Section III do not depend on a particular scattering mechanism, thus, without loss of generality, can be extended to any of the listed mechanisms.

To our knowledge the experimental techniques for investigating the thermoelectric transport in two-dimensional systems, such as graphene and twisted TMDs, are well-established (see, for instance, Refs.~~\cite{ghahari2016enhanced,zhang2023visualizing}). Here we anticipate that these methods can be used to explore the thermoelectric effects provided in our paper.

We wish to highlight another relevant system, so-called Weyl-Kondo semimetals, where our results can be directly applied~\cite{dzsaber2017kondo,chen2022topological, sur2024fully}. The Weyl-Kondo phase, which essentially occurs in strongly correlated systems, is characterized by a spontaneous Hall response. In these materials, T-symmetry remains unbroken, and Hall states arise in the absence of external magnetic fields due to the violation of inversion symmetry~\cite{dzsaber2021giant}. We believe that studying thermoelectric transport in such systems would be of great fundamental interest. 

\section{Acknowledement}

AVK acknowledges the support from Innovation Program for Quantum Science and Technology 2023ZD0300300. ZZA thanks the Ministry of Science and Higher Education of the Russian Federation (Goszadaniye) project No FSMG-2023-0011. EGI acknowledges financial support from the United Arab Emirates University under the Startup Grant No. G00004974.

\bibliography{apssamp}

\end{document}